\documentclass[12pt,amsmath,amssymb,aps,pra,superscriptaddress,notitlepage]{revtex4-1}
\usepackage{amsmath,amssymb,mathtools}
\usepackage{graphicx}
\usepackage{bm}
\usepackage[colorlinks, citecolor=blue, linkcolor=red]{hyperref}
\usepackage{braket}
\usepackage{xcolor, color, soul}

\begin{document}

\title{Superconductivity in single crystals of a quasi-one dimensional  infinite chain cuprate Sr$_x$Ca$_{1-x}$CuO$_2$ at 90 K}

\author{Neeraj K. Rajak}
\affiliation{School of Physics, IISER Thiruvananthapuram, Vithura, Thiruvananthapuram-695551, India}
\author{Dumpala Tirumalarao}
\affiliation{School of Physics, IISER Thiruvananthapuram, Vithura, Thiruvananthapuram-695551, India}
\author{Gourav Vaid}
\affiliation{School of Physics, IISER Thiruvananthapuram, Vithura, Thiruvananthapuram-695551, India}
\author{Sharath Kumar C}
\affiliation{School of Physics, IISER Thiruvananthapuram, Vithura, Thiruvananthapuram-695551, India}
\author{S. Athira}
\affiliation{School of Physics, IISER Thiruvananthapuram, Vithura, Thiruvananthapuram-695551, India}
\author{Govindarajan Prakash}
\affiliation{School of Physics, IISER Thiruvananthapuram, Vithura, Thiruvananthapuram-695551, India}
\author{Ashna Babu}
\affiliation{School of Physics, IISER Thiruvananthapuram, Vithura, Thiruvananthapuram-695551, India}
\author{Trupti Gaikwad}
\affiliation{School of Physics, IISER Thiruvananthapuram, Vithura, Thiruvananthapuram-695551, India}
\author{Shamili Chandradas}
\affiliation{Materials Science and Technology Division, CSIR-National Institute for Interdisciplinary Science and Technology, Thiruvananthapuram-
	695019, India}
\affiliation{Academy of Scientific and Innovative Research (AcSIR), CSIR-Human Resource Development Centre, (CSIR-HRDC) Campus, Postal Staff College Area, Sector 19, Kamla Nehru Nagar, Ghaziabad, Uttar-Pradesh-201002, India}
\author{Alex P. Andrews}
\affiliation{School of Physics, IISER Thiruvananthapuram, Vithura, Thiruvananthapuram-695551, India}
\author{Aneesh A.}
\affiliation{School of Physics, IISER Thiruvananthapuram, Vithura, Thiruvananthapuram-695551, India}
\author{Babu Varghese}
\affiliation{Sophisticated Analytical Instrumentation Facility, Indian Institute of Technology Madras, Chennai-600 036, India}
\author{Manoj Raama Varma}
\affiliation{Materials Science and Technology Division, CSIR-National Institute for Interdisciplinary Science and Technology, Thiruvananthapuram-
	695019, India}
\affiliation{Academy of Scientific and Innovative Research (AcSIR), CSIR-Human Resource Development Centre, (CSIR-HRDC) Campus, Postal Staff College Area, Sector 19, Kamla Nehru Nagar, Ghaziabad, Uttar-Pradesh-201002, India}
\author{Arumugam Thamizhavel}
\affiliation{Department of Condensed Matter Physics $\&$ Materials Science, Tata Institute of Fundamental Research, Mumbai, Maharashtra 400005, India}
\author{S. Ramakrishnan}
\affiliation{Department of Condensed Matter Physics $\&$ Materials Science, Tata Institute of Fundamental Research, Mumbai, Maharashtra 400005, India}
\author{D. Jaiswal-Nagar}
\email{deepshikha@iisertvm.ac.in}
\affiliation{School of Physics, IISER Thiruvananthapuram, Vithura, Thiruvananthapuram-695551, India}

\maketitle

\textbf{Although there is no complete theory of high temperature superconductivity, the importance of CuO$_2$ planes in cuprate superconductors is confirmed from both theory and experiments \cite{orenstein,keimer}. Strong Coulomb repulsion between electrons on the CuO$_2$ plane makes the resultant electron system highly correlated and a difficult problem to solve since exact solutions of many-body Hamiltonian in two dimensions do not exist. If however, superconductivity can arise in structures having chains rather than planes and having a high critical temperature, then the high temperature superconductivity problem could become more tractable since exact solutions in one dimension do exist \cite{giamarchi,essler}. In this paper, we report the observation of bulk superconductivity in single crystals of a cuprate Sr$_x$Ca$_{1-x}$CuO$_2$ at very high critical temperature, T$_c$, of $\sim$ 90 K whose structure reveals the presence of infinite double chains of Cu-O-Cu-O instead of CuO$_2$ planes, thus, ensuring quasi-one dimensional superconductivity. Bulk superconducting behaviour was observed in \textit{dc} magnetisation, \textit{ac} susceptibility as well as resistance measurements. The observation of bulk superconductivity in Sr$_x$Ca$_{1-x}$CuO$_2$ having chains of Cu-O-Cu-O rather than planes of CuO$_2$ at a high T$_c$ of 90 K is expected to profoundly impact our understanding of high temperature superconductivity.}\\ 
Cuprate high temperature superconductors (HTSCs) belong to a class of “strongly correlated electron systems” that comprise many unexplained phases and properties which have intrigued the condensed matter physics community for over three and a half decades \cite{orenstein,keimer} since the initial discovery in La$_{2-x}$Ba$_x$CuO$_4$ \cite{bednorz}. Superconductivity sets in the HTSCs by doping the parent Mott insulating antiferromagnetic phase suggesting that magnetism might be a common thread in understanding the microscopic origin of superconductivity in HTSCs \cite{anderson,dagatto}. Multipartite entanglement \cite{mathew} between the electrons in the strongly correlated system of a HTSC is expected to play a crucial role in understanding  high temperature superconductivity \cite{marel}. The majority of the cuprate HTSCs are characterised by infinite two-dimensional planes of CuO$_2$ that are linked to an apical oxygen, and there is strong evidence for an intimate link between the superconducting transition temperature T$_c$ and the apical oxygen-copper oxide distance \cite{pavarini,mahony}. So, superconductivity in systems in which such apical oxygens don’t exist, such as infinite layer superconductors (ILSs) \cite{li}, is quite intriguing. To understand HTSCs, exact solutions of many-body Hamiltonians such as the Hubbard or t-J models in two or three dimensions would be ideal. The non-existence of such exact solutions make the HTSC problem rather challenging. If, however, HTSCs could be found among quasi-one dimensional systems, it may facilitate our understanding of the HTSC problem since exact solutions of several many-body Hamiltonians exist in one dimension \cite{giamarchi,essler}. The d$_{x^2 - y^2}$ symmetry of the superconducting order parameter in HTSC already suggests that strong pairing of electrons for \textit{k} parallel to Cu-O-Cu bond direction would be favoured \cite{shen}. Quasi-one dimensional superconductivity has previously been reported in organic salts at rather low temperatures of $\sim$ 10 K and pressures of few kbars \cite{urayama}.\\ 
ILSs which have been proposed to be the parent compounds of all cuprate HTSCs can be thought of as the n $\rightarrow$ $\infty$ limit of a sequence of structures containing layers of CaCuO$_2$ where the n$^{th}$ layer of Ca$^{2+}$ is replaced by a charge reservoir layer (CRL) resulting in a chemical formula of the type [CRL](Ca)$_{n-1}$(CuO$_2$)$_n$ \cite{iyo}. The structure of such an infinite layer compound Ca$_{0.86}$Sr$_{0.14}$CuO$_2$ was first reported by Siegrist et al. to be tetragonal, with the space group \textit{P4/mmm} \cite{siegrist}.  However, this compound did not exhibit superconductivity. In fact, superconductivity at $\sim$ 110 K was reported in multi-phase powders of Sr$_x$Ca$_{1-x}$CuO$_2$ that were synthesized under a high pressure of few GPa and in an oxidizing environment of KClO$_4$ \cite{azuma,adachi,hiroi}. However, the superconducting transition with decreasing temperature was very broad, with small superconducting volume fractions indicative of impurities. Superconductivity in these multi-phase powders was found to be non-repeatable, with questions raised about contaminants from the crucible and oxidisers \cite{wang,cava}. By matching the peaks of the powder diffractogram of Sr$_x$Ca$_{1-x}$CuO$_2$ with those of IL material Ca$_{0.86}$Sr$_{0.14}$CuO$_2$, the structure of superconducting multi-phase powders of Sr$_x$Ca$_{1-x}$CuO$_2$ was also reported to be tetragonal \textit{P4/mmm}. However, there were several unindexed peaks in the powder diffractogram. Almost three decades after the initial reports of superconductivity in the multi-phase powders of Sr$_x$Ca$_{1-x}$CuO$_2$ that were found to have repeatability issues, in this paper we report bulk superconductivity in Sr$_x$Ca$_{1-x}$CuO$_2$ which has one dimensional Cu-O-Cu-O chain structure,  with a high, sharp transition temperature of $\sim$ 91 K. We expect that this will have far reaching consequences towards our understanding of high temperature superconductivity.

\begin{figure}
	\begin{center}
		\includegraphics[width=1\textwidth]{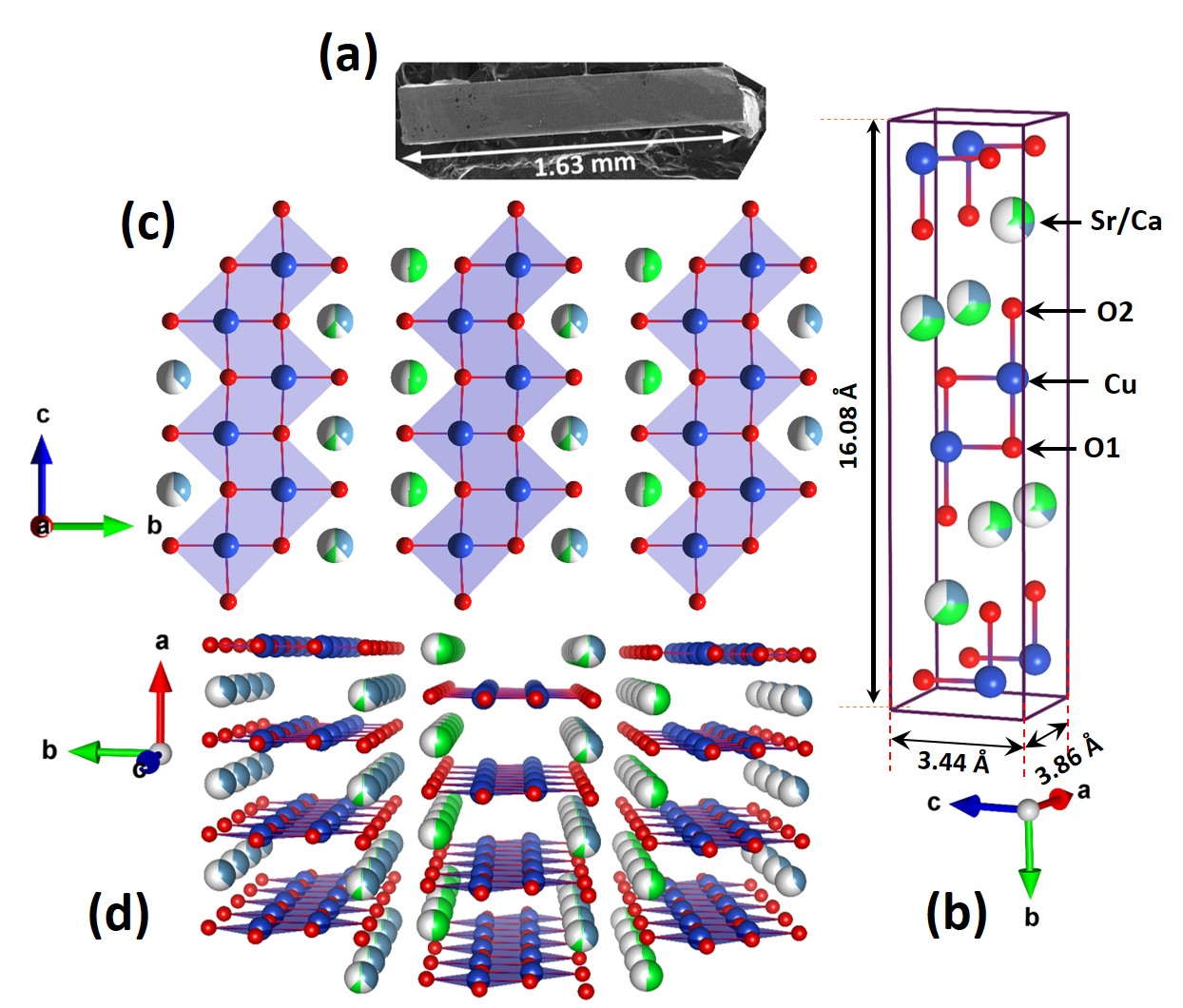}
		\caption{\textbf{Crystal structure of Sr$_x$Ca$_{1-x}$CuO$_2$}: (a) SEM image of a grown single crystal (b) Unit cell showing Cu-O-Cu-O chains separated by Sr/Ca atoms. Infinite double chains of Cu-O-Cu-O extending in the structure projected on (c) \textit{b}-\textit{c} plane and (d) \textit{a}-\textit{b} plane.}
		\label{fig:crystal structure}
	\end{center}
\end{figure}

The primary crystallization field of the HTSC Bi$_2$Sr$_2$CaCu$_2$O$_{8+x}$ (BSCCO) has a narrow temperature stability range from $\sim$ 900$^{\circ}$C to $\sim$ 825$^{\circ}$C and comprises a variety of other stable phases (apart from BSCCO), with Sr$_x$Ca$_{1-x}$CuO$_2$ being one of them \cite{styve}. Single crystals of Sr$_x$Ca$_{1-x}$CuO$_2$ were, in fact, obtained as a by-product of the self-flux technique of growing BSCCO single crystals under a small pressure of 0.234 N/cm$^2$ \cite{neeraj,neerajJAC} (Extended Data Fig. 1). Even though bismuth was not a part of the final structure, it was found that Bi$_2$O$_3$ as a starting component of the crystal growth of Sr$_x$Ca$_{1-x}$CuO$_2$ was absolutely necessary, suggesting that it acts as a flux in which the other components dissolve to initiate the crystal growth. It was also found that pressure application on the crucible during the crystal growth was necessary in the absence of which only crystals of BSCCO grow, suggesting that the vapour pressure exerted by the evaporating vapours of Bi$_2$O$_3$ in addition to the pressure exerted on the crucible lid helps to stabilize the Sr$_x$Ca$_{1-x}$CuO$_2$ phase. Phase equilibria studies by Roth et al. on the series Sr$_x$Ca$_{1-x}$CuO$_2$ revealed a very small stability range of \textit{x} = 0.15 $\pm$ 0.02 \cite{roth}. However, none of the compounds ranging from Sr$_{0.25}$Ca$_{0.75}$CuO$_2$ till Sr$_{0.5}$Ca$_{0.5}$CuO$_2$ were reported to be superconducting. In contrast, all our crystals having a minimum Sr content of 0.55 (Extended Data Table III) are superconducting. So, it seems that a critical amount of Sr$^{2+}$ of order 0.55 is necessary to induce superconductivity in this series. The crystals of Sr$_x$Ca$_{1-x}$CuO$_2$ grow as free-standing crystals amongst the BSCCO flux (Extended Data Fig. 1). The crystals are shiny and rectangular shaped with well-formed facets (see Fig. 1 (a)) unlike the layered morphology of BSCCO that makes BSCCO crystals break easily. In contrast, crystals of Sr$_x$Ca$_{1-x}$CuO$_2$ are found to be very hard and robust. From single crystal x-ray diffraction (ScXRD) measurements, the structure was found to be orthorhombic with space group \textit{CmCm}. Details of the refinement procedure are given in the Extended data. Crystallographic data for two crystals (named F and G) obtained by chipping a tiny portion of the crystals, are provided in Extended Data Table III. 

\begin{figure}
	\begin{center}
		\includegraphics[width=0.9\textwidth]{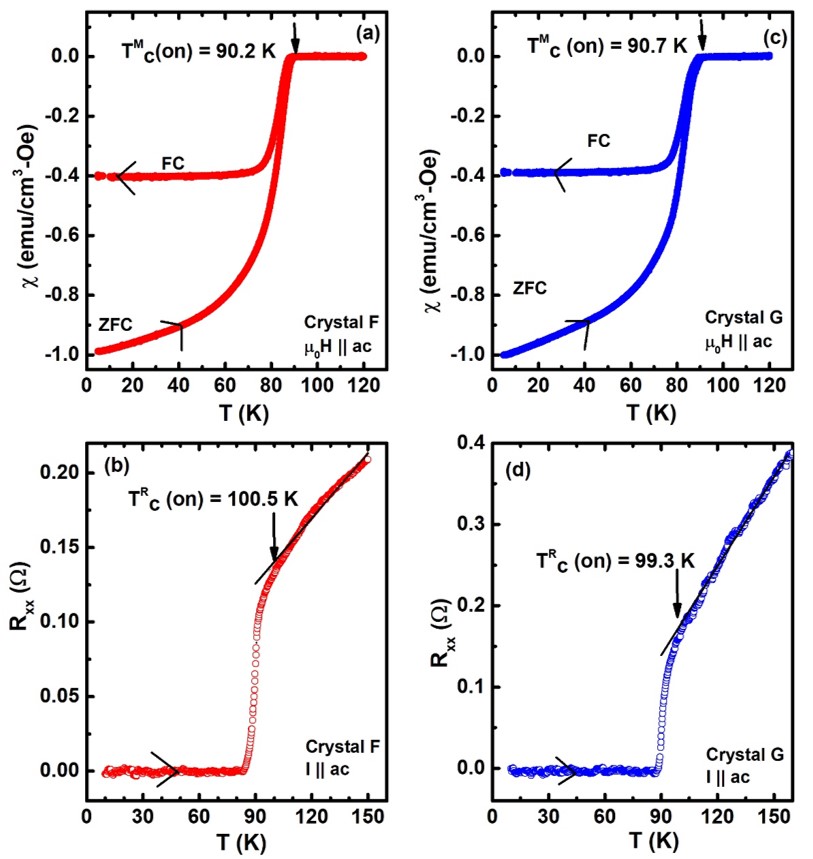}
		\caption{\textbf{Superconductivity in single crystals of Sr$_x$Ca$_{1-x}$CuO$_2$}: (a) and (c) represent \textit{dc} magnetic susceptibility measured on two different single crystals of Sr$_x$Ca$_{1-x}$CuO$_2$ labelled as F and G respectively. The measurements were made in the zero-field cooled (ZFC) as well as field-cooled (FC) mode at an applied field of 1 mT in the \textit{ac} plane of the crystals. Arrows depict the direction of measurement. Onset temperature of superconductivity, T$^M_c$(on) is marked in each panel. Demagnetisation factors have been taken into account while calculating the \textit{dc} susceptibility values. (b) and (d) represent the resistance R$_{xx}$ measured on crystals F and G respectively. The resistances were measured in the standard four probe geometry and the data collected while warming up. Superconducting onset temperature T$^R_c$(on) is depicted in each panel. T$^R_c$(on) in (b) and (d) denotes the temperature where the temperature dependence of resistance deviates from linearity (black straight lines in (b) and (d)) and starts to decrease.}
		\label{fig:Superconductivity}
	\end{center}
\end{figure}
      
The structure of Sr$_x$Ca$_{1-x}$CuO$_2$ is derived from SrCuO$_2$ with Ca$^{2+}$ replacing the Sr$^{2+}$ ions. From Fig. \ref{fig:crystal structure} (b), it can be seen that the structure comprises coplanar CuO$_4$ square plaquettes arranged in double parallel chains of CuO$_2$, one shifted with respect to the other by \textit{c}/2, and extending infinitely in the \textit{c}-direction. The Cu-O-Cu angle within a chain is close to 180$^{\circ}$ at 175.6$^{\circ}$, which would otherwise result in a strong antiferromagnetic correlation between the Cu$^{2+}$ ions \cite{goodenough}. The undoped parent compound SrCuO$_2$, is in fact, known to be a strongly correlated one dimensional spin 1$/$2 Heisenberg antiferromagnet \cite{zaliznyak,sologubenko}. The presence of magnetically inactive buffer layers of Sr/Ca in both the perpendicular directions of \textit{a} and \textit{b} (see Fig. \ref{fig:crystal structure} (c)) impart a strong one dimensional character to the Sr$_x$Ca$_{1-x}$CuO$_2$ system. Additionally, the 87.8$^{\circ}$ angle ($\sim$ 90$^{\circ}$) between Cu-O-Cu atoms of the double chain ensures that the coupling between the chains is extremely weak and frustrated \cite{goodenough}. The structure reported by Azuma et al. \cite{azuma} for the superconducting powders of Sr$_x$Ca$_{1-x}$CuO$_2$ was tetragonal with space group \textit{P4/mmm} based on the non-superconducting infinite layer structure of Ca$_{0.86}$Sr$_{0.14}$CuO$_2$ \cite{siegrist}. To confirm the orthorhombic structure of our superconducting crystals of Sr$_x$Ca$_{1-x}$CuO$_2$, apart from the ScXRD, we made a powder of the crystals and did a Rietveld refinement on the PXRD diffractogram with the structure parameters obtained from our ScXRD (Extended Data Fig. 4). The excellent fit with a low value of goodness of fit, GoF = 1.62, further confirms the orthorhombic structure of our Sr$_x$Ca$_{1-x}$CuO$_2$ crystals.\\ 
All our as-grown crystals of Sr$_x$Ca$_{1-x}$CuO$_2$ were found to be superconducting with \textit{x} varying between 0.64 to 0.55. Fig. \ref{fig:Superconductivity} shows the data for two single crystals with \textit{x} = 0.64 (Sr$_{0.64}$Ca$_{0.36}$CaCuO$_2$) and 0.55 (Sr$_{0.55}$Ca$_{0.45}$CaCuO$_2$) labelled as F and G respectively. Clear signatures of superconductivity were found in magnetization (Figs. \ref{fig:Superconductivity} (a) and (c)) as well as resistance (R) data (Figs. \ref{fig:Superconductivity} (b) and (d)). The critical temperature for the onset of superconductivity, T$^M_c$ (on), was denoted as the temperature where diamagnetism sets in the magnetization data. A high T$^M_c$ (on) of $\sim$ 90.2 K and 90.7 K was found in crystals F and G respectively. Bulk magnetic flux exclusion is found in crystals F and G with $\sim$ 100$\%$ superconducting volume fractions as indicated by the ZFC values in each crystal (c.f. Figs. \ref{fig:Superconductivity} (a) and (c)). The Meissner fractions for FC were correspondingly high, at $\sim$ 40$\%$ in both the crystals.\\
The resistive response of crystals F and G indicated that the normal state of these crystals is metal-like (dR/dT $>$ 0). T$^M_c$(on) is found to be $\sim$ 10 K smaller than T$^R_c$(on) (defined in the caption of Fig. \ref{fig:Superconductivity}) for crystals F and G, and corresponds well to the temperature at which dR/dT is maximal. Both the magnetic and resistive transitions were found to decrease with an increase in applied field as expected for a superconductor (Extended data Figs. 5 and 8) \cite{shang}. In the absence of high fields necessary to estimate the upper critical field H$_{c2}$, we marked the field at which resistance goes down to zero as an indicator of H$_{c2}$ \cite{shang}. For both the crystals, H$_{c2}$(T) was found to have a positive curvature at temperatures close to T$_c$(0), indicative of multiband superconductivity \cite{muller,gurevich}. Given the intimate correlation between the apical oxygen distance from the CuO$_2$ plane and T$_c$ that has been found in HTSCs of various kind \cite{pavarini,mahony}, a large T$_c$ of $\sim$ 91 K in a superconductor having a chain structure with no CuO$_2$ plane is fascinating! Since the crystal structure shows the presence of one dimensional double chains (see Fig. \ref{fig:crystal structure}), superconductivity in Sr$_x$Ca$_{1-x}$CuO$_2$ crystals implies quasi-one dimensional superconductivity arising from nearest neighbour Josephson coupling between the chains \cite{jaeferi}. Since Sr$^{2+}$ and Ca$^{2+}$ ions are similar sized, superconductivity in Sr$_x$Ca$_{1-x}$CuO$_2$ suggests self-doping.

\begin{figure}
	\centering
	\includegraphics[width=1\textwidth]{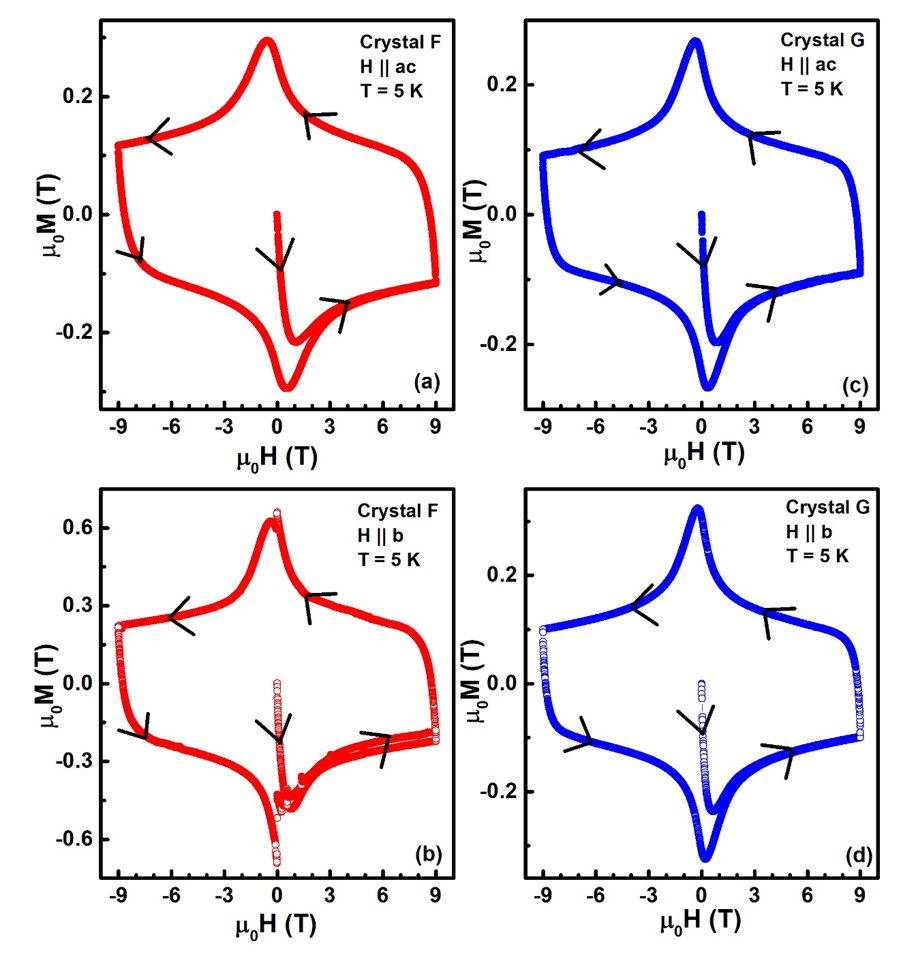}
	\caption{\textbf{Magnetisation hysteresis loops in Sr$_x$Ca$_{1-x}$CuO$_2$}: (a) and (c) depict the five quadrant magnetisation loop obtained by ramping the magnetic field upto $\pm$9 T in crystals F and G respectively, with the applied field, $\mu_0$H, parallel to \textit{ac} and measured at a temperature of 5 K. (b) and (d) represent the corresponding magnetisation loops in crystals F and G respectively for $\mu_0$H, parallel to \textit{b}. Arrows in each panel mark the direction of field sweep. The curves were recorded using the ZFC protocol. }
	\label{fig:Hysterisis loop}
\end{figure}
  
Field dependent magnetization-M(H) loops, measured at a temperature of 5 K are shown in Fig. \ref{fig:Hysterisis loop} for both the crystals. The curves display the usual response of a Type-II superconductor \cite{jaiswal-nagar}, further confirming superconductivity in these infinite chain crystals. The field of full penetration, H$_p$ (5 K), was found to have very large values in the Tesla range, quite unlike those observed in other HTSC’s \cite{cohen,manju}. Additionally, the M(H) response was not found to have much anisotropy with similar values of magnetization obtained for $\mu_0$H $||$ \textit{ac} or $\mu_0$H $||$ \textit{b}. Furthermore, for $\mu_0$H $||$ b, a flux jump was observed in crystal F in the third and the fifth quadrant at fields lower than H$_p$, indicative of a possible symmetry reorientation transition in the vortex lattice \cite{jaiswal-nagar}. The temperature dependent M-H data recorded for the two crystals showed a shift in the lower critical field, H$_{c1}$, to lower fields with an increase in temperature (Extended Data Fig. 6). The resultant H$_{c1}$-T curves estimated in a limited temperature range are also shown in Extended Data Fig. 6.

\begin{figure}
	\centering
	\includegraphics[width=1\textwidth]{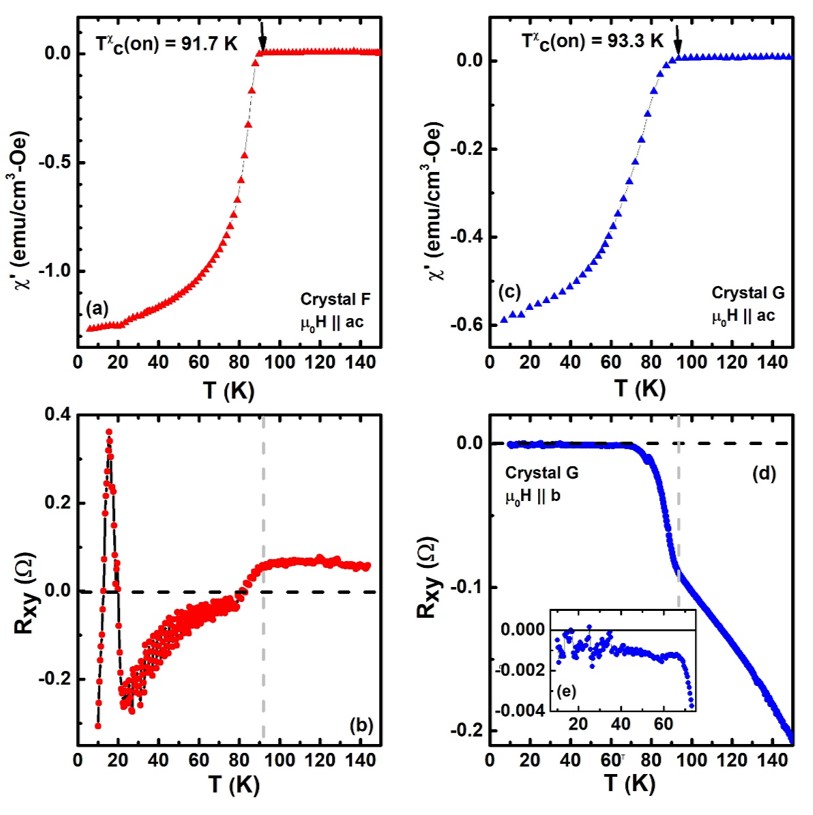}
	\caption{\textbf{\textit{ac} susceptibility and Hall measurements in Sr$_x$Ca$_{1-x}$CuO$_2$}: Temperature dependent  \textit{ac} susceptibility measurements done on (a) crystal F and (c) crystal G in an applied field B of 0.03 mT parallel to \textit{ac} measured in the ZFC protocol. Transition temperature is marked in each graph. (b) and (d)  show the temperature variation of Hall resistance measured on crystals F and G respectively, in an applied field of 0.6 T such that I $||$ \textit{ac} and $\mu_0$H $||$ \textit{b}. The measurements were made in the warm-up mode. Inset (e) shows the low temperature Hall resistance, R$_{xy}$, of crystal G on an expanded scale.}
	\label{fig:acHall}
\end{figure}

Temperature dependent \textit{ac} susceptibility measurements performed on the two crystals also exhibited bulk superconductivity as seen in Figs. \ref{fig:acHall} (a) and (c). The onset temperature for superconductivity, T$_c^\chi$(on), is found to be consistent but slightly higher than that found from \textit{dc} susceptibility measurements (c.f. Fig. \ref{fig:Superconductivity}). The superconducting volume fraction of crystal F was $\sim$ 100$\%$, similar to that obtained in \textit{dc} susceptibility measurements; however, that of crystal G was lower, at $\sim$ 60$\%$.\\
To understand the nature of the charge carriers in the normal state of the superconductor, we performed Hall effect measurements-a direct probe of the Fermi surface topology and charge carrier density \cite{leboeuf,badoux}. From the temperature variation of R$_{xy}$ (Figs. \ref{fig:acHall} (b) and (d)) and consequently the Hall coefficient R$_H$ = \textit{t}*R$_{xy}$/\textit{B}, where \textit{t} is the thickness of the crystals (Extended Data Fig. 9), it was found that the Hall coefficient is positive in the normal state of crystal F but is negative in crystal G. So, either the charge carriers in the normal state of crystal F are holes or the Fermi surface has a positive curvature \cite{leboeuf,badoux}. Equivalently, either the charge carriers in crystal G are negative or its Fermi surface has a negative curvature.\\
The Hall resistance of crystal G retains its negative sign in the superconducting state till the lowest measured temperature (see the inset Figs. \ref{fig:acHall} (e)). However, in crystal F, R$_{xy}$ changes its sign thrice on entering the superconducting state at T $\sim$ 81 K, 21 K and 13 K. A single sign change of R$_{xy}$ has been observed in underdoped crystals of YBa$_2$Cu$_3$O6$_{6+\delta}$ \cite{leboeuf,badoux} and La$_{2-x}$Ba$_x$CuO$_4$ \cite{adachiLBCO}, a double sign change observed in atomically flat Bi$_2$Sr$_2$CaCuO$_{8+x}$ \cite{zhao} while no sign change is seen in underdoped and overdoped crystals of La$_{2-x}$Sr$_x$CuO$_4$ \cite{ando}. However, to our knowledge such a triple sign change in the superconducting state of any high T$_c$ superconductor has not been reported previously and maybe an indicator of either coexisting electron and hole pockets in the Fermi surface or reconstruction of the Fermi surface itself. The possibility of the mobility of the charge carriers having different temperature dependencies is not ruled out.\\ 
    
To conclude, bulk superconductivity is observed in single crystals of Sr$_x$Ca$_{1-x}$CuO$_2$ though \textit{dc} magnetization, resistance and \textit{ac} susceptibility measurements. Crystal structure indicates the presence of double infinite chains along the \textit{c}-axis of the crystals, indicating quasi-one dimensional superconductivity in the system. This structure is very different from the one proposed by Azuma et al. \cite{azuma} who derived it from the non-superconducting tetragonal structure of Ca$_{0.86}$Sr$_{0.14}$CuO$_2$. Our study calls for a reexamination of superconductivity in those HTSC compounds that have Cu-O-Cu-O chains in their structure like YBa$_2$Cu$_3$O6$_{6+\delta}$ and YBa$_2$Cu$_4$O$_{8-\delta}$ having single and double chain structures respectively. The d$_{x^2 - y^2}$ symmetry of superconducting order parameter in HTSCs already suggests that the electrons pair strongly along the \textit{k}-direction parallel to Cu-O bonds but are not paired perpendicular to it \cite{shen}. So, the discovery of superconductivity in a system with infinite Cu-O-Cu chains but no CuO$_2$ planes, and having a high critical temperature of 90 K is expected to have far-reaching consequences towards the understanding of the problem of high T$_c$ superconductivity.     
\section*{Methods Summary}
Single crystal XRD was measured on Bruker’s Kappa APEX II CCD diffractometer equipped with graphite monochromatized Mo-K$_\alpha$ radiation. Magnetisation measurements were performed on the VSM attachment of Quantum Design’s physical property measurement system (PPMS, Model Evercool-II) in the zero field cooled (ZFC) mode where the system was cooled to the lowest temperature in the absence of a field after which a field was applied and data taken while warming up. In the field cooled (FC) mode, the data was taken while cooling down the sample in an applied field. The standard terminology of 5 quadrant M-H measurements in a superconductor is used: 0 $\rightarrow$ H$_{max}$-1$^{st}$ quadrant; H$_{max}$ $\rightarrow$ 0-2$^{nd}$ quadrant; 0 $\rightarrow$ -H$_{max}$-3$^{rd}$ quadrant; H$_{max}$ $\rightarrow$ 0-4$^{th}$ quadrant and 0 $\rightarrow$ H$_{max}$ (second time)-5$^{th}$ quadrant. Resistance measurements were performed on the electrical transport option (ETO) option of the 14 T PPMS. The contacts for the resistance measurements were made in the standard four-probe geometry using silver paint. 
\section*{Author Contribution} 
D.J.N. designed the study, N.K.R, D.T, G.V. and T.G. grew the single crystals. A.P.A. and B.V. solved the crystal structure. N.K.R, D.T., G.V., S.R.C.,S.A.,A.B.,P.G.,T.G.,
S.,A.P.A.,A.A.,M.R.,A.T.,S.R. and D.J.N. performed the magnetisation and transport measurements. D.J.N. wrote the manuscript with inputs from all co-authors. 
\begin{acknowledgments}
The authors thank G. Baskaran, H. R. Krishnamurthy, Arun K. Grover, A. Sundaresan and Siddhartha Lal for critically examining the manuscript and giving very valuable feedbacks. D. J-N. acknowledges financial support from SERB, DST, Govt. of India (Grants No. YSS/2015/001743 and CRG/2021/001262).
\end{acknowledgments}

\nocite{*}

\bibliography{}


\end{document}


\title{Superconductivity in single crystals of a quasi-one dimensional  infinite chain cuprate Sr$_x$Ca$_{1-x}$CuO$_2$ at 90 K}
	
\author{Neeraj K. Rajak}
\affiliation{School of Physics, IISER Thiruvananthapuram, Vithura, Thiruvananthapuram-695551, India}
\author{Dumpala Tirumalarao}
\affiliation{School of Physics, IISER Thiruvananthapuram, Vithura, Thiruvananthapuram-695551, India}
\author{Gourav Vaid}
\affiliation{School of Physics, IISER Thiruvananthapuram, Vithura, Thiruvananthapuram-695551, India}
\author{Sharath Kumar C}
\affiliation{School of Physics, IISER Thiruvananthapuram, Vithura, Thiruvananthapuram-695551, India}
\author{S. Athira}
\affiliation{School of Physics, IISER Thiruvananthapuram, Vithura, Thiruvananthapuram-695551, India}
\author{Govindarajan Prakash}
\affiliation{School of Physics, IISER Thiruvananthapuram, Vithura, Thiruvananthapuram-695551, India}
\author{Ashna Babu}
\affiliation{School of Physics, IISER Thiruvananthapuram, Vithura, Thiruvananthapuram-695551, India}
\author{Trupti Gaikwad}
\affiliation{School of Physics, IISER Thiruvananthapuram, Vithura, Thiruvananthapuram-695551, India}
\author{Shamili Chandradas}
\affiliation{Materials Science and Technology Division, CSIR-National Institute for Interdisciplinary Science and Technology, Thiruvananthapuram-
	695019, India}
\affiliation{Academy of Scientific and Innovative Research (AcSIR), CSIR-Human Resource Development Centre, (CSIR-HRDC) Campus, Postal Staff College Area, Sector 19, Kamla Nehru Nagar, Ghaziabad, Uttar-Pradesh-201002, India}
\author{Alex P. Andrews}
\affiliation{School of Physics, IISER Thiruvananthapuram, Vithura, Thiruvananthapuram-695551, India}
\author{Aneesh A.}
\affiliation{School of Physics, IISER Thiruvananthapuram, Vithura, Thiruvananthapuram-695551, India}
\author{Babu Varghese}
\affiliation{Sophisticated Analytical Instrumentation Facility, Indian Institute of Technology Madras, Chennai-600 036, India}
\author{Manoj Raama Varma}
\affiliation{Materials Science and Technology Division, CSIR-National Institute for Interdisciplinary Science and Technology, Thiruvananthapuram-
	695019, India}
\affiliation{Academy of Scientific and Innovative Research (AcSIR), CSIR-Human Resource Development Centre, (CSIR-HRDC) Campus, Postal Staff College Area, Sector 19, Kamla Nehru Nagar, Ghaziabad, Uttar-Pradesh-201002, India}
\author{Arumugam Thamizhavel}
\affiliation{Department of Condensed Matter Physics $\%$ Materials Science, Tata Institute of Fundamental Research, Mumbai, Maharashtra 400005, India}
\author{D. Jaiswal-Nagar}
\email{deepshikha@iisertvm.ac.in}
\affiliation{School of Physics, IISER Thiruvananthapuram, Vithura, Thiruvananthapuram-695551, India}

\begin{abstract}
In the Extended Data, we provide details of (I) single crystal growth of Sr$_x$Ca$_{1-x}$CuO$_2$, (II) EDX measurements on Sr$_x$Ca$_{1-x}$CuO$_2$, (III) crystal structure, (IV) Laue diffraction, (V) Rietveld refinement, (VI) magnetisation and (VII) resistance measurements on Sr$_x$Ca$_{1-x}$CuO$_2$ to confirm various facets of the superconductivity in these materials.
\end{abstract}

\maketitle

\section{Crystal growth}
\begin{figure}[hbtp]
	\centering
	\includegraphics[scale=0.51]{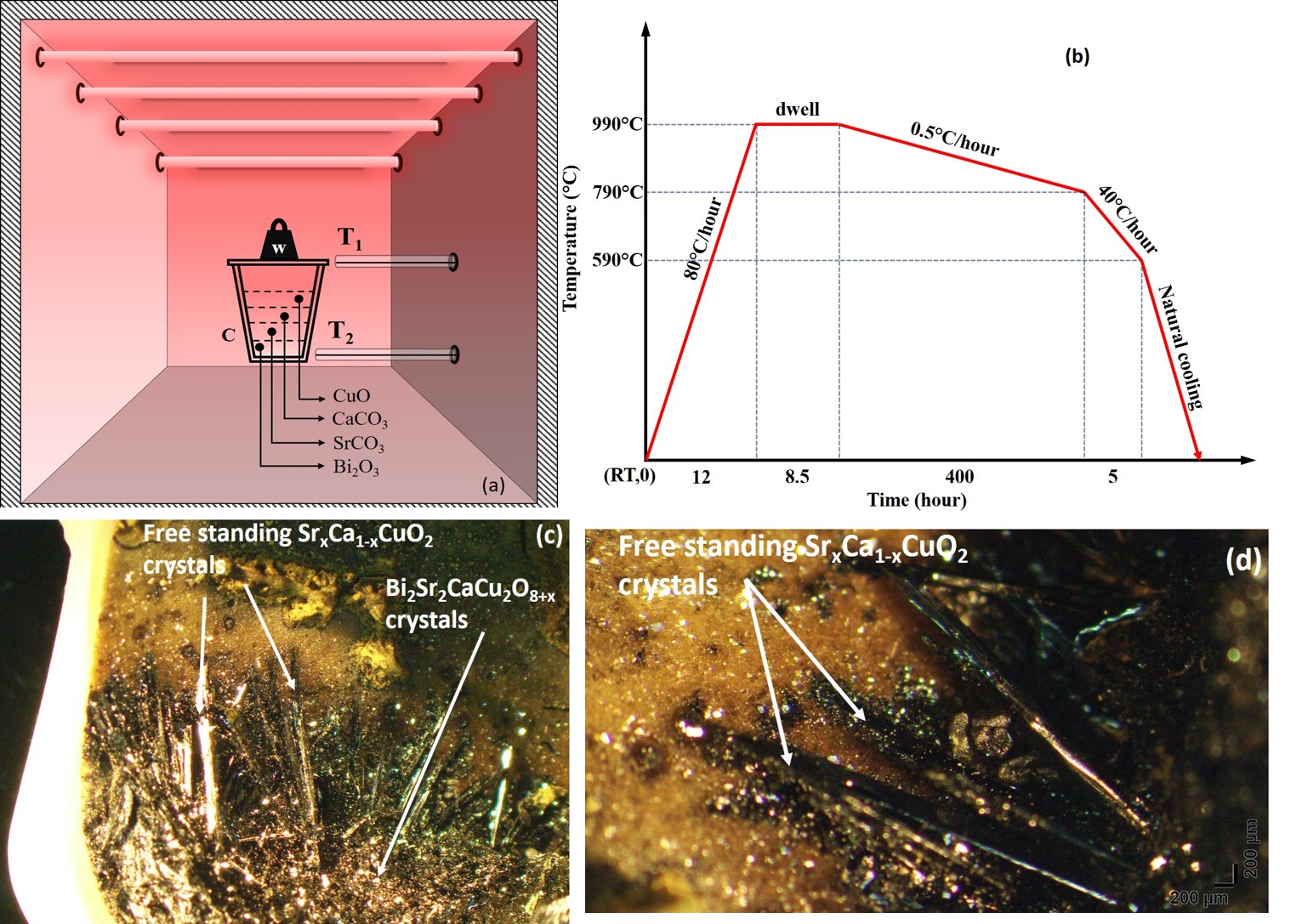}
	\caption{\textbf{Crystal growth and optical images of Sr$_x$Ca$_{1-x}$CuO$_2$}: (a) Powders of Bi$_2$O$_3$, SrCO$_3$, CaCO$_3$ and CuO placed layer wise in the crucible with a weight W placed on top. T$_1$ and T$_2$ are the thermocouples that measure the temperatures at the top and bottom of the crucible. (b) Temperature-time profile showing segments of time and temperature sweep rates to which the crucible shown in (a) was subjected. (c) Image of Sr$_x$Ca$_{1-x}$CuO$_2$ single crystals grown in an alumina crucible. The image was taken with an optical microscope (Model SZX10) that shows Sr$_x$Ca$_{1-x}$CuO$_2$ crystals growing as free standing crystals. The background black mass consists of Bi$_2$Sr$_2$CaCu$_2$O$_{8+x}$ single crystals. (d) Zoomed in image of (c) to show Sr$_x$Ca$_{1-x}$CuO$_2$ crystals more clearly. They can be seen as having a well defined morphology of a cuboid.}
	\label{fig:crystal growth}
\end{figure}

Single crystals of Sr$_x$Ca$_{1-x}$CuO$_2$ were obtained as a by-product of the self-flux technique of growing BSCCO single crystals under a small pressure of 0.234N/cm$^2$ that was applied on the alumina crucible while growing the crystals \cite{neeraj,neerajJAC}. For the crystal growth, powders of Bi$_2$O$_3$ (99.999$\%$), SrCO$_3$ (99.999$\%$), CaCO$_3$ (99.999$\%$) and CuO (99.999$\%$), obtained from Merch-Aldrich, were weighed in the molar ratio of Bi:Sr:Ca:Cu = 2:2:1:2. In this crystal growth, the powders were not ground and kept in a 50 mL alumina crucible layer-by-layer. Since Bi$_2$O$_3$ has the lowest melting point, it was kept at the bottom of the crucible, followed by SrCO$_3$, CaCuO$_3$ and CuO as shown in Fig. \ref{fig:crystal growth} (a). Bi$_2$O$_3$ acts as the flux in which the other reactants dissolve to initiate the crystal growth. A lid was then, put on the top of the alumina crucible and weights corresponding to the pressure of 0.234N/cm$^2$ were put on the lid. The crucible was then subjected to the temperature profile as shown in Fig. \ref{fig:crystal growth} (b) to inititate the crystal growth. 

It was found that the crystal growth of Sr$_x$Ca$_{1-x}$CuO$_2$ is extremely sensitive to the applied pressure. Even a variation of pressure of $\pm$0.02 N/cm$^2$ results in only the growth of BSCCO crystals with no Sr$_x$Ca$_{1-x}$CuO$_2$ crystals. With an applied pressure of 0.234N/cm$^2$, we have been only able to grow Sr$_x$Ca$_{1-x}$CuO$_2$ crystals having \textit{Sr} stoichiometry between 0.55-0.6. The crystals grow as free standing crystals in the background of Bi$_2$Sr$_2$CaCu$_2$O$_{8+x}$ (BSCCO) crystals as shown in Fig. \ref{fig:crystal growth} (c).

\section{EDX Measurements}

\begin{figure}[hbtp]
	\centering
	\includegraphics[scale=0.5]{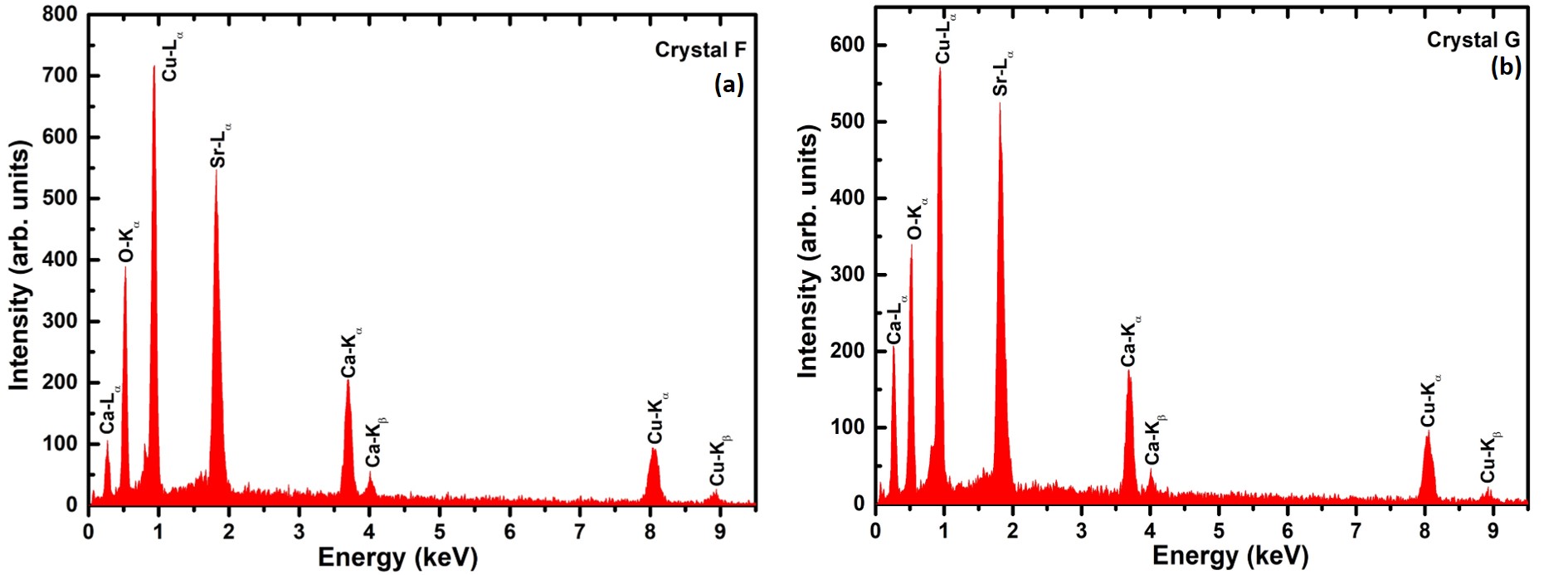}
	\caption{\textbf{EDX spectra obtained on crystals F and G}: Energy variation of intensity corresponding to (a) crystal F and (b) crystal G. Strong peaks corresponding to Cu, O, Ca and Sr can be seen. No Bi peak was observed.}
	\label{fig:EDX}
\end{figure}

The initial stoichiometry was obtained by doing an energy dispersive x-ray analysis (EDX) on the grown crystals. Since the crystals were obtained with using Bi$_2$O$_3$ as flux, it was quite a possibility that Bi was part of the stoichiometry. However, it was found that Bi was either very small or negligible in the obtained crystals. We also realised that the small amount of Bi seen in some spectra could be due to the surrounding flux which was BSCCO-2212 itself. So, we cleaned all the crystals with emery paper of grade 1200 before doing an EDX measurement. EDX measurements were carried out on a scanning electron microscope (Model No. FEI Nova NanoSEM 450) fitted with an EDX attachment. The stoichiometry Sr$_x$Ca$_{1-x}$CuO$_2$ of the crystals grown in different batches over many years was confirmed by statistically measuring the spectra on a given crystal at different locations in the crystal. Figures \ref{fig:EDX} (a) and (b) show the obtained EDX spectra on crystals F and G respectively measured at an accelerating voltage of 15 kV. From the two spectra, it is clearly seen that the element bismuth is not present in either of the crystals within the resolution of the measuring apparatus.\\ 
Tables \ref{tableF} and \ref{tableG} show the relative stoichiometry of crystals F and G respectively. From the table, the relative ratio of Sr+Ca : Cu is obtained as 0.95:1 for crystal F and 0.96:1 for crystal G. The individual mole ratio for each element is found to be Sr:Ca:Cu = 0.51: 0.46:1 in crystal F and Sr:Ca:Cu = 0.54: 0.42:1 in crystal G. 

\begin{table}[hbtp]
	\centering
	\includegraphics[scale=0.8]{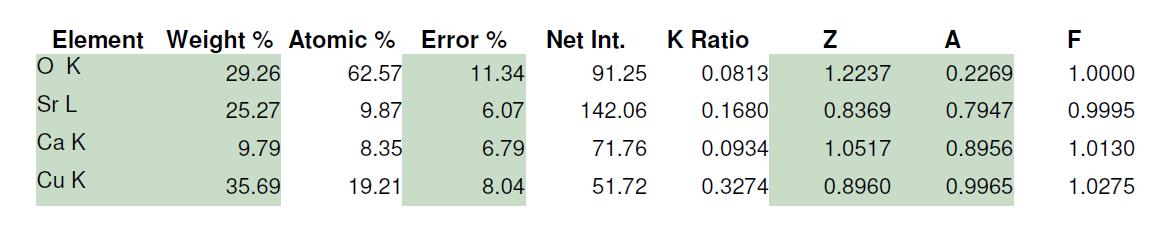}
	\caption{Atomic percentage of the elements corresponding to crystal F}
	\label{tableF}
\end{table}

\begin{table}[hbtp]
	\centering
	\includegraphics[scale=0.8]{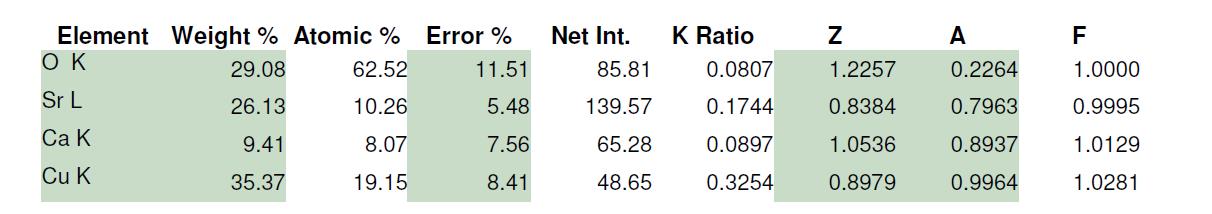}
	\caption{Atomic percentage of the elements corresponding to crystal G}
	\label{tableG}
\end{table}

\section{Crystal structure}
To accurately estimate the empirical formula as well as to get the structure of the grown crystals correctly, we performed single crystal x-ray diffraction (ScXRD) on crystals F and G. The measurements were done at a temperature of 296(2) K on Bruker’s Kappa APEX II CCD diffractometer equipped with graphite monochromatized Mo-K$_{\alpha}$ radiation having a wavelength of 0.71073 \AA. The intensity data was collected using $\omega$ and $\phi$ scans with frame width of 0.5$^{\circ}$. Frame integration and data reduction were performed using Bruker's SAINT/XPREP software \cite{bruker}. Multi-scan absorption corrections were applied to the data using SADABS (Bruker 2016) program \cite{sadabs}. The structures were solved using the software SHELXT-2014/5 \cite{shelxt} and refined using SHELXL-2018/3 \cite{shelxl}. Structure solution gave common position for both strontium and calcium atoms. The position was treated as a disorder site with partial occupancies for Sr and Ca. The structures were refined with occupancies of Sr and Ca as additional refinement variables with sum of the occupancies restrained as 0.25 which is the full occupancy for the site. Also, the thermal parameters of Sr and Ca were restrained to be equal. Some isotropic restraints were applied to thermal parameters of all atoms as these parameters have very small values due to the rigid nature of the structure. As the atomic numbers of calcium (20) and copper (29) are sufficiently apart and intensity data quality quite good, the occupancy refinement is expected to yield correct value for Sr/Ca stoichiometry. Details of the refinement for crystals F and G are given below in Table \ref{table:crystallographic parameters} below. From the Table \ref{table:crystallographic parameters}, it is clear that the lattice parameters of Sr$_x$Ca$_{1-x}$CuO$_2$ are widely separated from that of BSCCO-2212 \cite{neerajJAC}.

\begin{table}[htb!]
	\centering
	\begin{tabular}{|c |c |c |} 
		\hline
		Property & Crystal F	& Crystal G	\\
		\hline 
		Dimensions & 0.038 x 0.035 x 0.020 mm$^3$	& 0.068 x 0.048 x 0.038 mm$^3$	\\
		\hline
		Formula	& Sr 0.64 Ca0.36 Cu O2	& Sr0.55 Ca0.45 Cu O2	\\
		\hline
		Formula Weight	& 166.16	& 161.53	\\
		\hline
		Crystal System	& Orthorhombic	& Orthorhombic	\\
		\hline
		Space group	& CmCm	& CmCm	\\
		\hline
		& \textit{a} = 3.4731(3)\AA ~~~ $\alpha = 90^{\circ}$ & \textit{a} = 3.4428(5)\AA ~~ $\alpha = 90^{\circ}$\\
		Unit cell dimensions & \textit{b} = 16.1600(14)\AA ~~ $\beta = 90^{\circ}$ & \textit{b} = 16.086(2)\AA ~~ $\beta = 90^{\circ}$\\ 
		& \textit{c} = 3.8769(3)\AA	~~~ $\gamma = 90^{\circ}$ & \textit{c} = 3.8638(5)\AA ~~ $\gamma = 90^{\circ}$\\
		\hline
		Volume & 217.59(3) \AA$^3$	& 213.98(5)  \AA$^3$	\\
		\hline
		Density (calculated) & 5.072 Mg/m$^3$	& 5.014 Mg/m$^3$	\\
		\hline
		Absorption coefficient	& 26.080 mm$^{-1}$	& 24.350 mm$^{-1}$	\\
		\hline
		$\theta$ range	& 5.046 to 28.294$^{\circ}$	& 5.070 to 28.421$^{\circ}$	\\
		\hline
		& -4$<$=h$<$=4, 	& -4$<$=h$<$=4, \\
		Index ranges   & -21$<$=k$<$=21, & -21$<$=k$<$=13, \\
		
		& -5$<$=l$<$=5	& -4$<$=l$<$=5	\\
		\hline
		Reflections collected	& 3032 & 725 \\
		\hline
		Independent reflections	& 176 [R(int) = 0.0361] & 175 [R(int) = 0.0240] \\
		\hline
		Max. and min. transmission	& 0.51 and 0.28	& 0.52 and 0.31	\\
		\hline
		Goodness-of-fit on F$^2$	& 1.287	& 1.181	\\
		\hline
		Final R indices & R1 = 0.0156, 	& R1 = 0.0188, 	\\
		(I$>$2sigma(I))	& wR2 = 0.0396	& wR2 = 0.0472	\\
		\hline
		R indices (all data) & R1 = 0.0157,	& R1 = 0.0205,	 \\
		& wR2 = 0.0396	& wR2 = 0.0478 \\
		\hline
		Extinction coefficient	& 0.0205(19)	& 0.013(2)	\\
		\hline
		Largest diff. peak and hole	& 0.489 and -0.505e.\AA$^{-3}$	& 0.660 and -0.923e.\AA$^{-3}$	\\
		\hline
	\end{tabular}
	\caption{Summary of crystallographic information of Sr$_x$Ca$_{1-x}$CuO$_2$ measured on crystals F and G}
	\label{table:crystallographic parameters}
\end{table}

\section{Laue Diffraction:}
\begin{figure}[h!]
	\centering
	\includegraphics[scale=0.6]{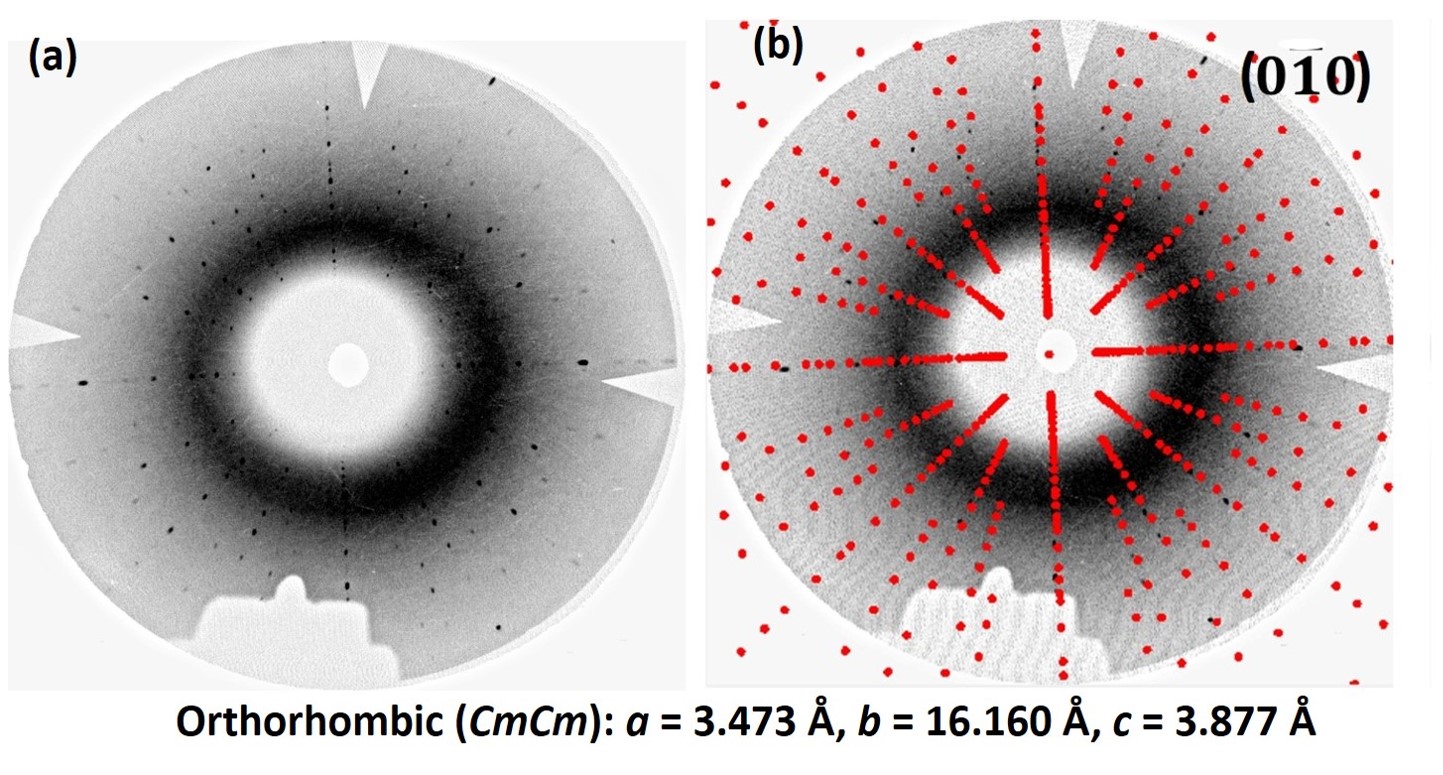}
	\caption{\textbf{Laue diffraction pattern obtained on crystal F}: (a) Experimentally observed X-ray back-Laue pattern on crystal F with the surface parallel to the growth direction. (b) Simulated pattern using Orient Express software showing alignment along (010).}
	\label{fig:Laue}
\end{figure}

To reconfirm the structure obtained from single crystal X-ray diffraction as well as to find the relevant crystallographic direction with respect to the crystal morphology, we resorted to back-reflection Laue diffraction that was done on crystal F. The zone axis was aligned parallel to the thickness of the crystal. The observed Laue pattern, shown in Fig. \ref{fig:Laue} (a) was analysed with a simulated pattern, shown in Fig. \ref{fig:Laue} (b). The simulated pattern was generated using Orient Express software \cite{orient} by inputting the orthorhombic structure with lattice constants \textit{a} = 3.473 \AA, \textit{b} = 16.160 \AA and \textit{c} = 3.877 \AA. The simulated pattern overlap the experimentally obtained diffraction pattern for the (010) plane. So, the thickness of the crystal is along \textit{b} - (010). Since the cross-sectional area of the other two dimensions was very small, we couldn’t obtain the Laue pattern for the other two directions. So, for magnetisation measurements, we aligned the crystals either parallel to thickness (parallel to \textit{b}) or parallel to length (parallel to \textit{ac} plane).

\section{Rietveld refinement}

\begin{figure}[h!]
	\centering
	\includegraphics[scale=0.6]{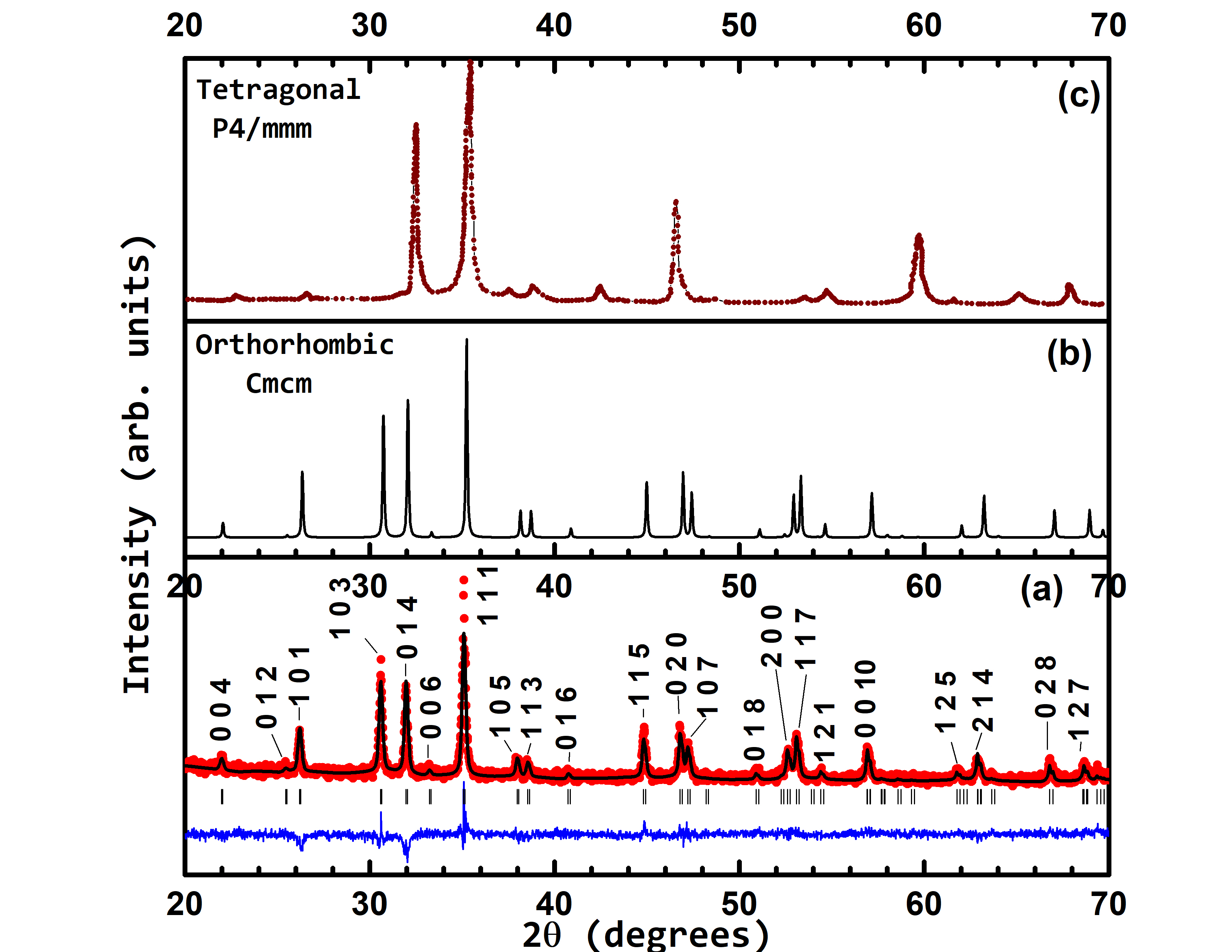}
	\caption{\textbf{Powder diffraction pattern on Sr$_x$Ca$_{1-x}$CuO$_2$}: (a) Red filled circles denote experimental PXRD obtained on crushed single crystal of Sr$_x$Ca$_{1-x}$CuO$_2$ while the black curve corresponds to Rietveld refined fit obtained by incorporating background correction, preferred orientation correction, absorption correction etc. Blue curve is the difference curve while green vertical lines indicate Bragg peaks. (b) PXRD generated from single crystal XRD CmCm structure. (c) PXRD of tetragonal Sr$_x$Ca$_{1-x}$CuO$_2$ extracted from \cite{azuma}.}
	\label{fig:Rietveld}
\end{figure}

To confirm the orthorhombic crystal structure of Sr$_x$Ca$_{1-x}$CuO$_2$ obtained by single crystal XRD, we resorted to Rietveld refinement, a powerful technique of obtaining structure, lattice modulation, quantitative phase analysis of competing and co-crystallising phases etc. based on a powder diffractogram \cite{dusek,rietveld,petricek,neerajJAC}. Rietveld refinement also confirms the structure over an entire crystal since the whole crystal is made into a powder for getting the powder diffractogram. So, a single crystal of Sr$_x$Ca$_{1-x}$CuO$_2$ was crushed into a powder and a powder diffractogram obtained on it as shown in Fig. \ref{fig:Rietveld} (a). The powder obtained by crushing a single crystal of Sr$_x$Ca$_{1-x}$CuO$_2$ was put on a PANalytical zero background sample holder (model number SI SUBSTER 32MM) and pressed gently with a dried glass slide. The powder diffractogram was recorded on a PANalytical powder diffractometer in Bragg-Brentano geometry with a copper source, having both Cu K$_{\alpha1}$ and Cu K$_{\alpha2}$ radiation, corresponding to wavelengths of 1.540 \AA and 1.544 \AA, respectively. The counts were recorded for 2 hours in the 2$\theta$ range of 20-70$^{\circ}$ in steps of 0.016$^{\circ}$. Rietveld refinement of the PXRD data was done using the JANA2006 program \cite{petricek}.

\begin{figure}[h!]
	\centering
	\includegraphics[scale=1]{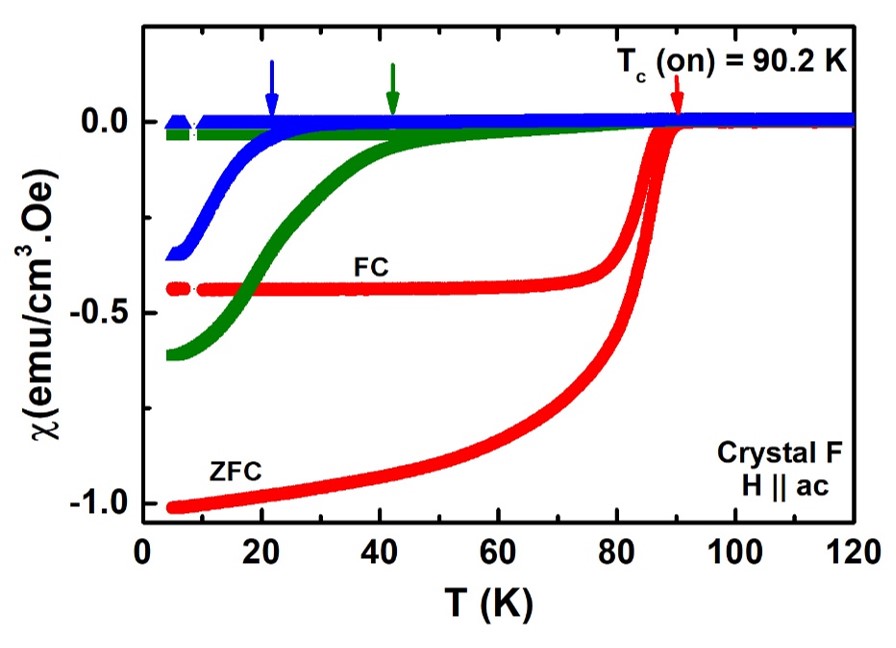}
	\caption{\textbf{Decrease of T$_c$ with an increase in the applied field in crystal F}: Red filled circles denote the \textit{dc} susceptibility obtained in crystal F at an applied field of 1 mT while olive green filled squares and blue filled triangles denote the \textit{dc} susceptibility at applied fields of 0.1 T and 0.5 T respectively. Measurements were done in both ZFC as well as FC modes of measurement.}
	\label{fig:TcshiftH}
\end{figure}

To compare this diffractogram with that of the tetragonal structure obtained by Azuma \cite{azuma}, the data from \cite{azuma} was extracted and plotted in Fig. \ref{fig:Rietveld} (c). PXRD graph generated from ScXRD is shown in Fig. \ref{fig:Rietveld} (b). It can be clearly seen that the powder x-ray diffractogram of an orthorhombic structure has additional peaks at 30.8$^{\circ}$, 45$^{\circ}$, 51.5$^{\circ}$, 52.8$^{\circ}$, 57.2$^{\circ}$, 63.4$^{\circ}$ and 67.2$^{\circ}$. Additionally, the difference between a tetragonal structure and an orthorhombic structure is clearly manifested by the splitting of the peak at 46.5$^{\circ}$ into two peaks at 47$^{\circ}$ and 47.2$^{\circ}$ \cite{mizuguchi,manjuybco}.  Comparing Figs. \ref{fig:Rietveld} (a) and (b), it can be clearly seen that the experimentally obtained PXRD matches perfectly well with that of the simulated spectra. When compared to the PXRD of the tetragonal structure obtained by Azuma \cite{azuma}, the presence of the additional peaks mentioned above as well as the peak split at 46$^{\circ}$ in Fig. \ref{fig:Rietveld} (a) confirms the orthorhombic structure of our Sr$_x$Ca$_{1-x}$CuO$_2$ crystals. Rietveld refined curve reproduces the experimentally obtained PXRD data very well (black curve in Fig. \ref{fig:Rietveld} (a)) confirming the orthorhombic \textit{CmCm} structure of our superconducting Sr$_x$Ca$_{1-x}$CuO$_2$ crystals. Final refined parameters are tabulated in Table \ref{table:Rietveld}. Goodness of fit for the final refinement was obtained as 1.62 indicating extremely good matching of the simulated structure and the obtained powder diffractogram \cite{neerajJAC}. 

\begin{table}[h]
	\centering
	\hspace{-1.5cm}
	\begin{tabular}{|c|c|c|c|c|c|c|c|c|}
		\hline 
		Label   &Atom& Wyckoff & x   &   y     &   z  & B$_{iso}$ & Site  & Occ  \\ \hline 
		
		Sr1 & Sr & 4c &0.5 & 0.8307  & 0.75  & 0.0036  &m2m   & 0.62  \\ \hline 
		Ca1 & Ca & 4c &0.5 & 0.8270  & 0.75  & 0.0036  &m2m   & 0.38  \\ \hline 
		Cu1 & Cu & 4c &0.5 & 0.5614  & 0.75  & 0.003167&m2m   & 1     \\ \hline  
		O1  & O  & 4c &0.5 & 0.6800  & 0.75  & 0.006067&m2m   & 1     \\ \hline 
		O2  & O  & 4c &0.5 & 0.4434  & 0.75  & 0.005333&m2m   & 1     \\ \hline   	
	\end{tabular}
	\caption{Atomic positions, Wyckoff, site symmetry, occupancy and the isothermal vibration parameter of the Cmcm phase.}
	\label{table:Rietveld}
\end{table} 

\section{Magnetisation Measurements}

\begin{figure}[h!]
	\centering
	\includegraphics[scale=0.8]{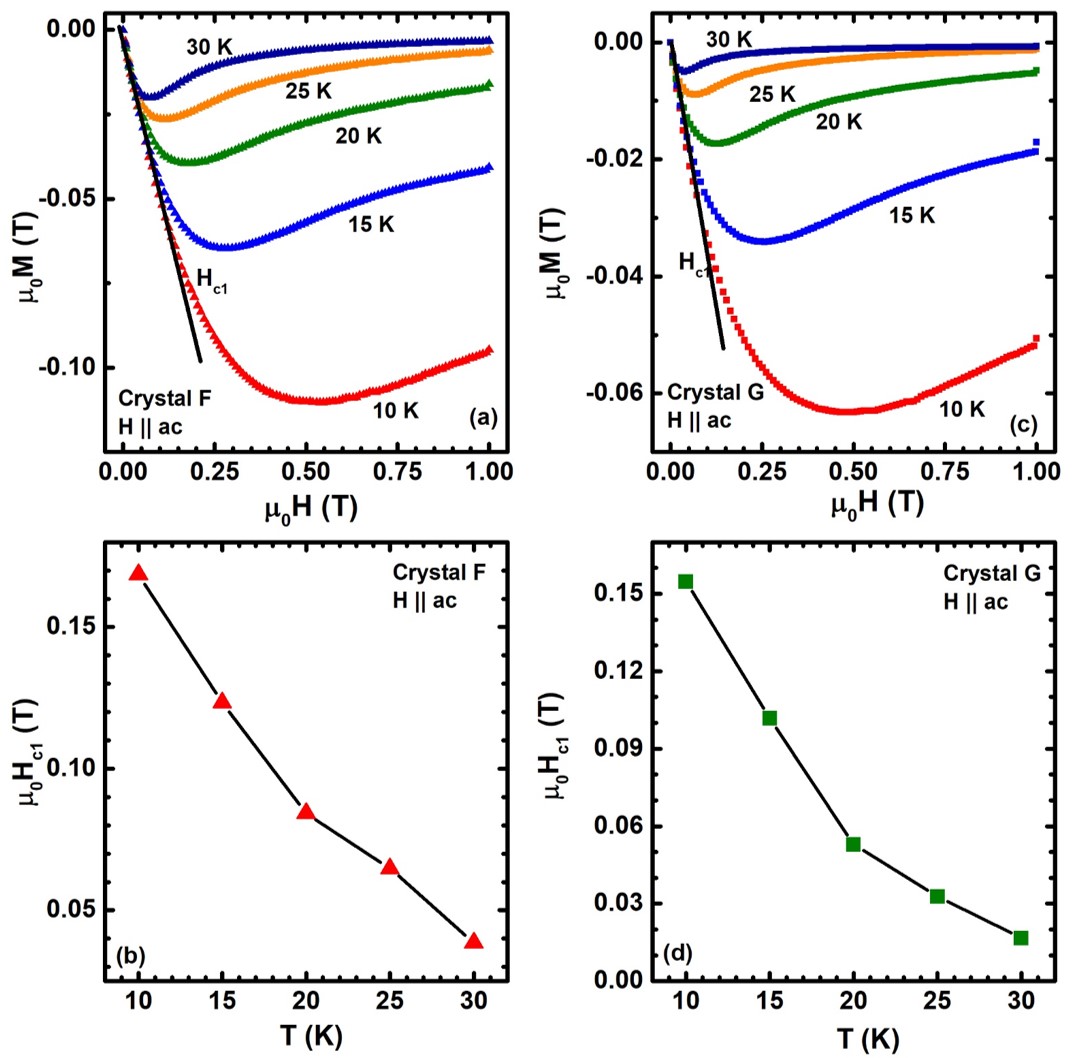}
	\caption{\textbf{H$_{c1}$-T obtained from virgin magnetisation loops of Sr$_x$Ca$_{1-x}$CuO$_2$}: Temperature variation of first quadrant virgin M-H loops in crystals (a) F and (c) G. Straight line fits on each M-H loop is shown by black solid line. H$_{c1}$ in marked as deviation from linearity in each panel. Filled red triangles and filled green squares denote the temperature variation of lower critical field H$_{c1}$ in crystals (b) F and (d) G respectively.}
	\label{fig:Hc1}
\end{figure}
The superconducting transition temperature was found to shift to lower temperatures with an increase in applied magnetic fields as expected for a superconductor \cite{song} and shown in the magnetisation response of crystal F for $\mu_0$H $||$ \textit{ac} in Fig. \ref{fig:TcshiftH}. As the applied field increases from 1 mT to 0.1 T to 0.5 T, the superconducting T$_c$ depresses from 90.2 K to 42.9 K to 22.4 K.\\ 
The M-H curves were similarly found to shift with temperature as expected in a superconductor. In order to estimate the temperature variation of the lower critical field, H$_{c1}$, we measured the virgin loops in crystals F and G at a few temperatures as shown in Figs. \ref{fig:Hc1} (a) and (b) respectively. It can be seen that the magnetisation of the virgin curves is linear till the lower critical field H$_{c1}$ above which it starts to deviate from linearity. Using this criterion, H$_{c1}$-T was made for few temperatures as shown in Figs. \ref{fig:Hc1} (b) and (d) for crystals F and G respectively.

\begin{figure}[h!]
	\centering
	\includegraphics[scale=0.8]{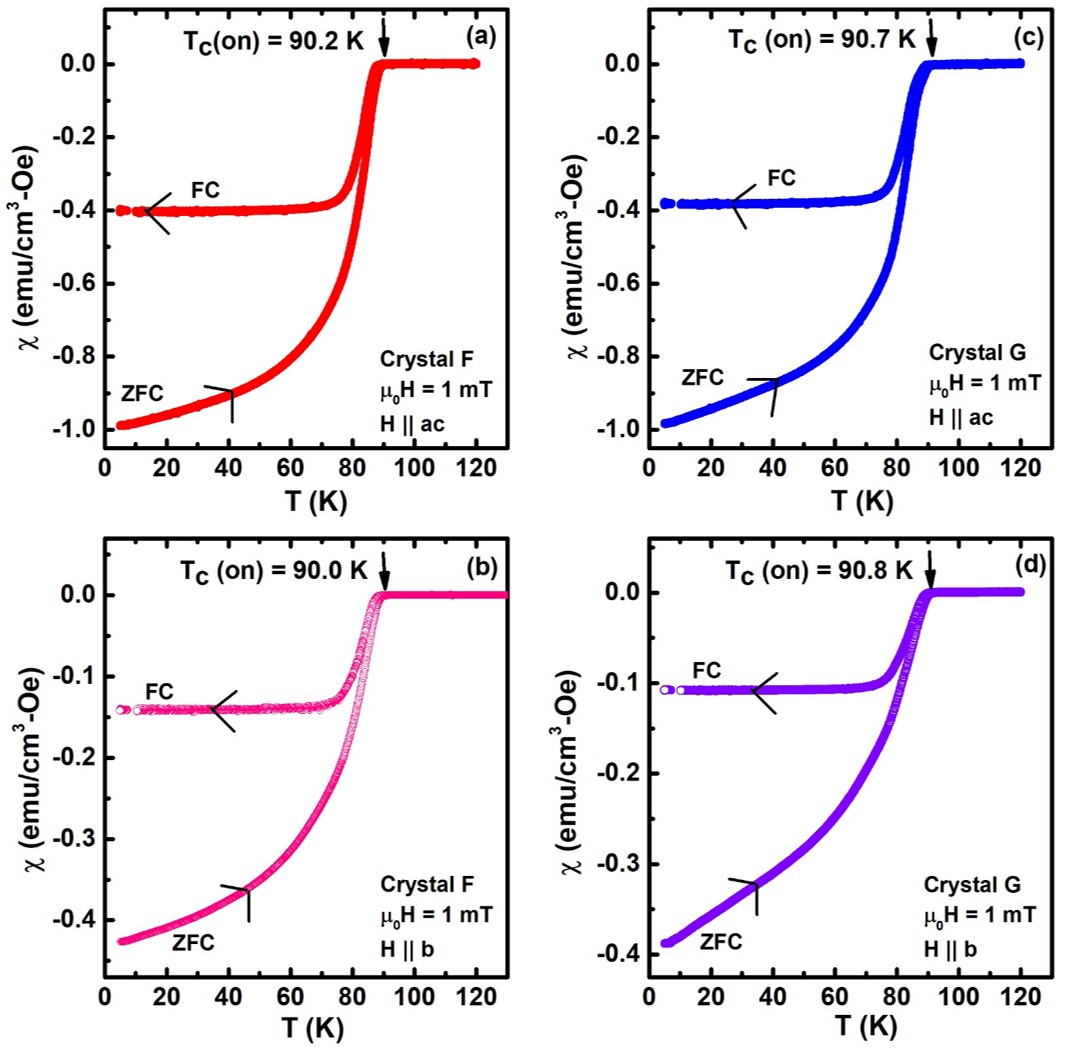}
	\caption{\textbf{Orientation dependent magnetisation measurements to check anisotropy}: Magnetisation response obtained for applied field $\mu_0$H $||$ \textit{ac} in crystal F (a) and crystal G (c). (b) and (d) depict the magnetisation response for $\mu_0$H $||$ \textit{b} in crystals F and G respectively. Superconducting T$_c$ is marked in each.}
	\label{fig:Orientation}
\end{figure} 
It was found that the superconducting T$_c$ is insensitive to the crystal orientation as seen in Fig. \ref{fig:Orientation}. Figs. \ref{fig:Orientation} (a) and (c) depict the magnetisation response of crystals F and G respectively for $\mu_0$H $||$ \textit{ac} while Figs. \ref{fig:Orientation} (b) and (d) depict the same response for $\mu_0$H $||$ \textit{b}. As can be seen from the figures, the T$_c$(on) has approximately the same values for both the crystals irrespective of their orientation relative to the magnetic field, although the Meissner fractions are smaller for $\mu_0$ H $||$ b.
\section{Resistance Measurements}
\begin{figure}[h!]
	\centering
	\includegraphics[scale=0.8]{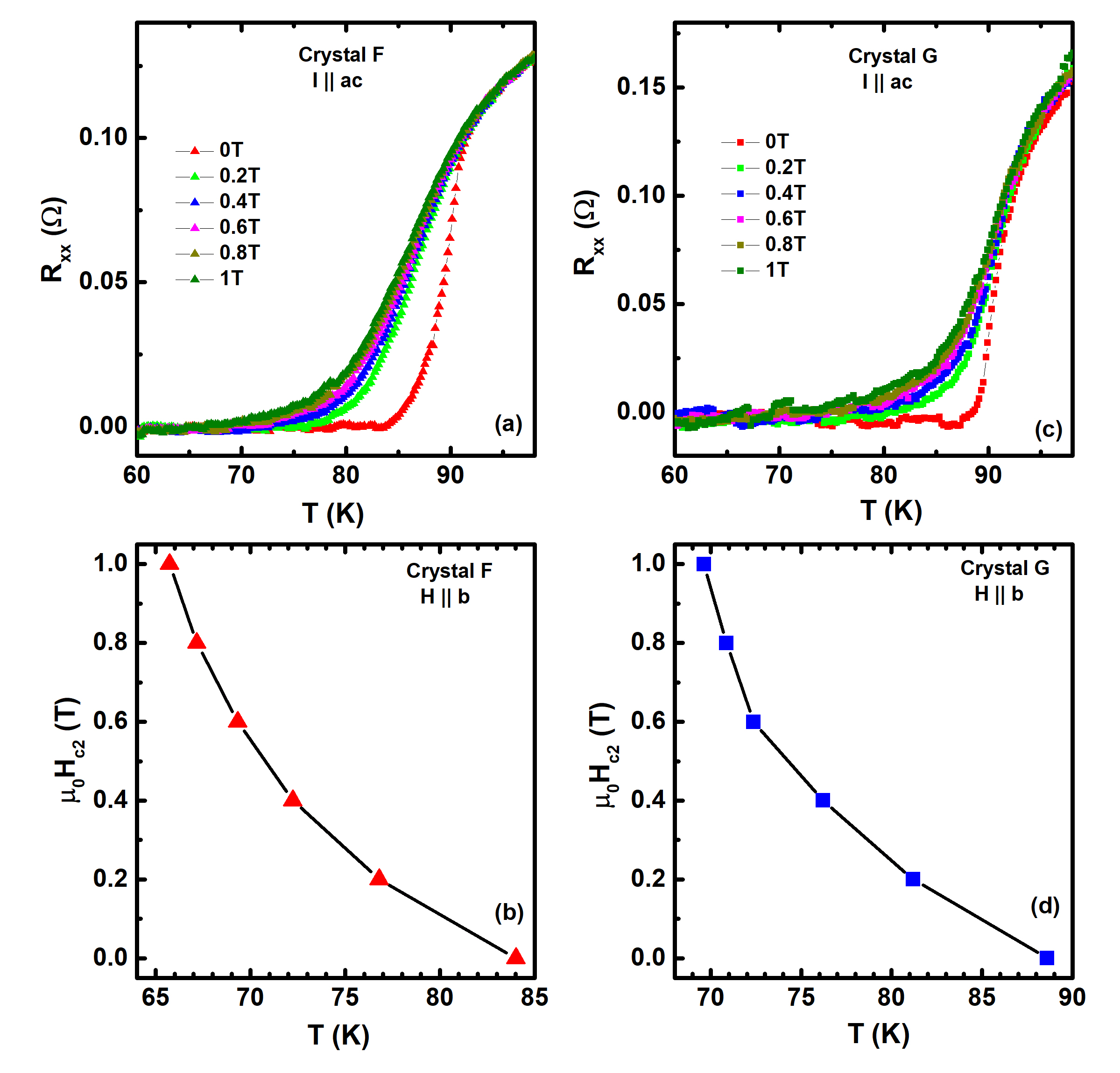}
	\caption{\textbf{T$_c$ shift with field and H$_{c2}$-T near T$_c$(0) obtained from field dependent R-T measurements}: (a) and (c) depict the temperature dependent resistance as a function of applied field ($\mu_0$H $||$ \textit{b}) for crystals F and G respectively. (b) and (d) show the H$_{c2}$-T phase diagrams for crystals F and G respectively.}
	\label{fig:Hc2}
\end{figure}
To further confirm the superconductivity in these infinite chain cuprates, we repeated the resistance measurements on the crystals in the presence of an applied magnetic field  and found the superconducting T$_c$ to shift to lower values with an increase in the value of applied field. Figs. \ref{fig:Hc2} (a) and (b) show the temperature dependent resistance measurements done on crystals F and G respectively with field parallel to \textit{b}. From the figure, it can be clearly seen that as the applied field is increased from 0 T to 1 T, the T$_c$ shifts to lower values systematically, similar to the shift in superconducting T$_c$ observed in the magnetisation data above. This observation confirms the presence of bulk superconductivity in these infinite chain crystals.
\begin{figure}[h!]
	\centering
	\includegraphics[scale=0.8]{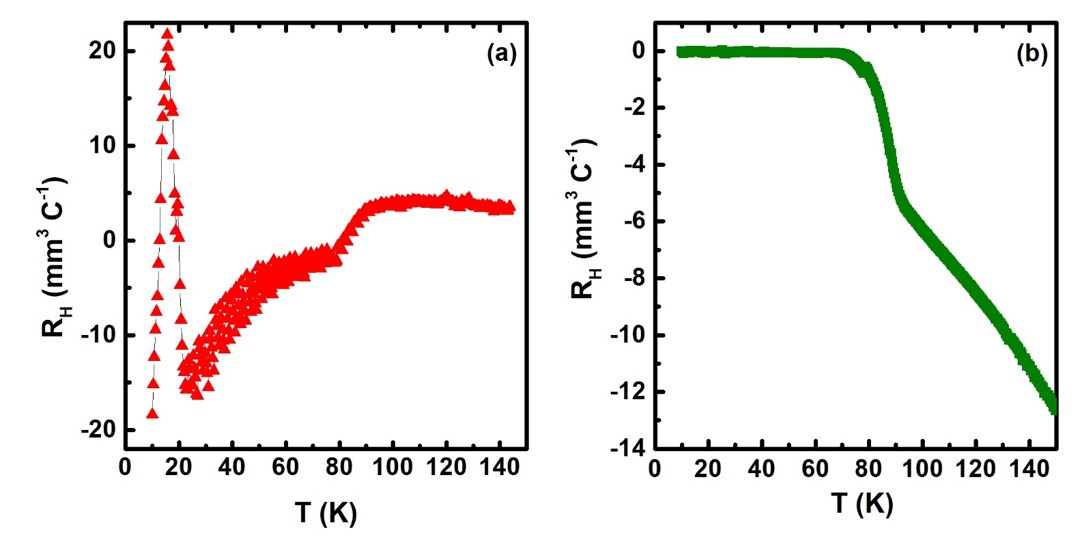}
	\caption{\textbf{Hall coefficient in crystals of Sr$_x$Ca$_{1-x}$CuO$_2$}: (a) Temperature variation of Hall coefficient for crystals (a) F and (b) G respectively.}
	\label{fig:Hall coefficient}
\end{figure}  

In absence of high field measurements, the values of fields at which the resistance goes to zero in the R-T measurements are taken as indicators of H$_{c2}$ values \cite{shang}. The temperature variation of H$_{c2}$ near T$_c$(0) is shown in Figs. \ref{fig:Hc2} (b) and (d) for crystals F and G respectively where the curvature of $\mu_0$H$_{c2}$(T) is positive indicating multiband superconductivity \cite{gurevich}. However, more experiments need to be done to confirm this.\\
The Hall coefficient R$_H$ = \textit{t}*R$_{xy}$/\textit{B}, where \textit{t} is the thickness of the crystals and \textit{B} is the applied magnetic field, was calculated from measurements of the transverse resistance in the presence of a magnetic field for the crystals F and G. The thickness of crystals F and G was measured to be 360 $\mu$m and 367 $\mu$m respectively. The temperature variation of the Hall coefficient is shown in Fig. \ref{fig:Hall coefficient}.

\bibliography{}